\documentclass{article}
\usepackage{spconf,amsmath,graphicx}
\usepackage{multirow}
\usepackage{dblfloatfix}
\usepackage[pagebackref=true,breaklinks=true,letterpaper=true,colorlinks,bookmarks=false]{hyperref}
\usepackage{enumitem}
\setlist{nosep, leftmargin=14pt}

\usepackage{mwe} 

\title{A Registration- and Uncertainty-Based Framework for White Matter Tract Segmentation with Only One Annotated Subject}
\address{
$^{1}$School of Computer Science, University of Sydney, Australia\\
$^{2}$Brigham and Women’s Hospital, Harvard Medical School, USA}

\begin{document}
%
\maketitle

\begin{abstract}
White matter (WM) tract segmentation based on diffusion magnetic resonance imaging (dMRI) plays an important role in the analysis of human health and brain diseases. However, the annotation of WM tracts is time-consuming and needs experienced neuroanatomists. In this study, to explore tract segmentation in the challenging setting of minimal annotations, we propose a novel framework utilizing only one annotated subject (subject-level one-shot) for tract segmentation. Our method is constructed by proposed registration-based peak augmentation (RPA) and uncertainty-based refining (URe) modules. RPA module synthesizes pseudo subjects and their corresponding labels to improve the tract segmentation performance. The proposed URe module alleviates the negative influence of the low-confidence voxels on pseudo subjects. Experimental results show that our method outperforms other state-of-the-art methods by a large margin, and our proposed modules are effective. Overall, our method achieves accurate whole-brain tract segmentation with only one annotated subject. Our code is available at \url{https://github.com/HaoXu0507/ISBI2023-One-Shot-WM-Tract-Segmentation}. 

\end{abstract}
\section{Introduction}
\label{sec:1}
Diffusion magnetic resonance imaging (dMRI) \cite{Basser1994-ti} is the only non-invasive method for in-vivo mapping of the human brain white matter (WM). A WM tract is a set of white matter fibers (axons) forming a corticocortical or cortico-subcortical connection in the brain \cite{Zhang2022-sj}. WM tract segmentation based on dMRI is important for analyzing WM characteristics in healthy and diseased brains \cite{Zhang2022-sj,Wasserthal2018-jj}. 

Recently, deep-learning-based tract segmentation methods have been widely used to achieve outstanding segmentation accuracy \cite{Wasserthal2018-jj,Zhang2020-vm,Xue2022-mz,xue2023-102759,Astolfi2020-zk,Li2020-gn}. These deep-learning methods usually train a network using a large-scale annotated dataset. For example, TractSeg \cite{Wasserthal2018-jj} utilizes a U-Net structure \cite{Ronneberger2015-eu} to segment WM tracts using fiber orientation distribution function (fODF) peaks by training and validating on over 80 annotated subjects. However, obtaining annotations for WM tracts is time-consuming and needs experienced neuroanatomists. Several semi-supervised WM tract segmentation methods have been proposed, such as tract-level few/one-shot \cite{Lu2022-iv,Liu2022-se} and few-shot segmentation methods using limited annotated subjects \cite{Lu2021-fp}.
\cite{Lu2022-iv,Liu2022-se} transfer tract segmentation knowledge from fully-supervised tracts into few/one-shot annotated tracts. Although tract-level few/one-shot tract segmentation implemented in \cite{Lu2022-iv,Liu2022-se} has successfully transferred knowledge to tracts with insufficient annotations, a large number of subjects still need to be annotated. \cite{Lu2021-fp} designs two pretext tasks to enable tract segmentation with a few annotated subjects. However, to our knowledge, no method has explored WM tract segmentation under the extremely minimal annotation condition.
That is, only one subject is annotated in the dataset for training, denoted as subject-level one-shot learning. Exploring this minimal annotation setting can be helpful for clinical applications due to the difficulty of tract annotations.

In this study, we propose a novel deep learning framework for subject-level one-shot tract segmentation, which leverages proposed registration-based peak augmentation (RPA) and uncertainty-based refining (URe) modules. Our contributions are as follows: 1) We propose an effective deep learning framework, achieving accurate whole-brain tract segmentation results with only one annotated subject; 2) To improve the segmentation performance under the scarcity of annotated subjects, we synthesize pseudo peak subjects and their corresponding tract segmentation labels through RPA module;
and 3) To further improve model accuracy, we propose a URe module to facilitate the self-supervised learning process by refining synthesized pseudo labels. 

\begin{figure*}[t]

\centering\includegraphics[width=15cm]{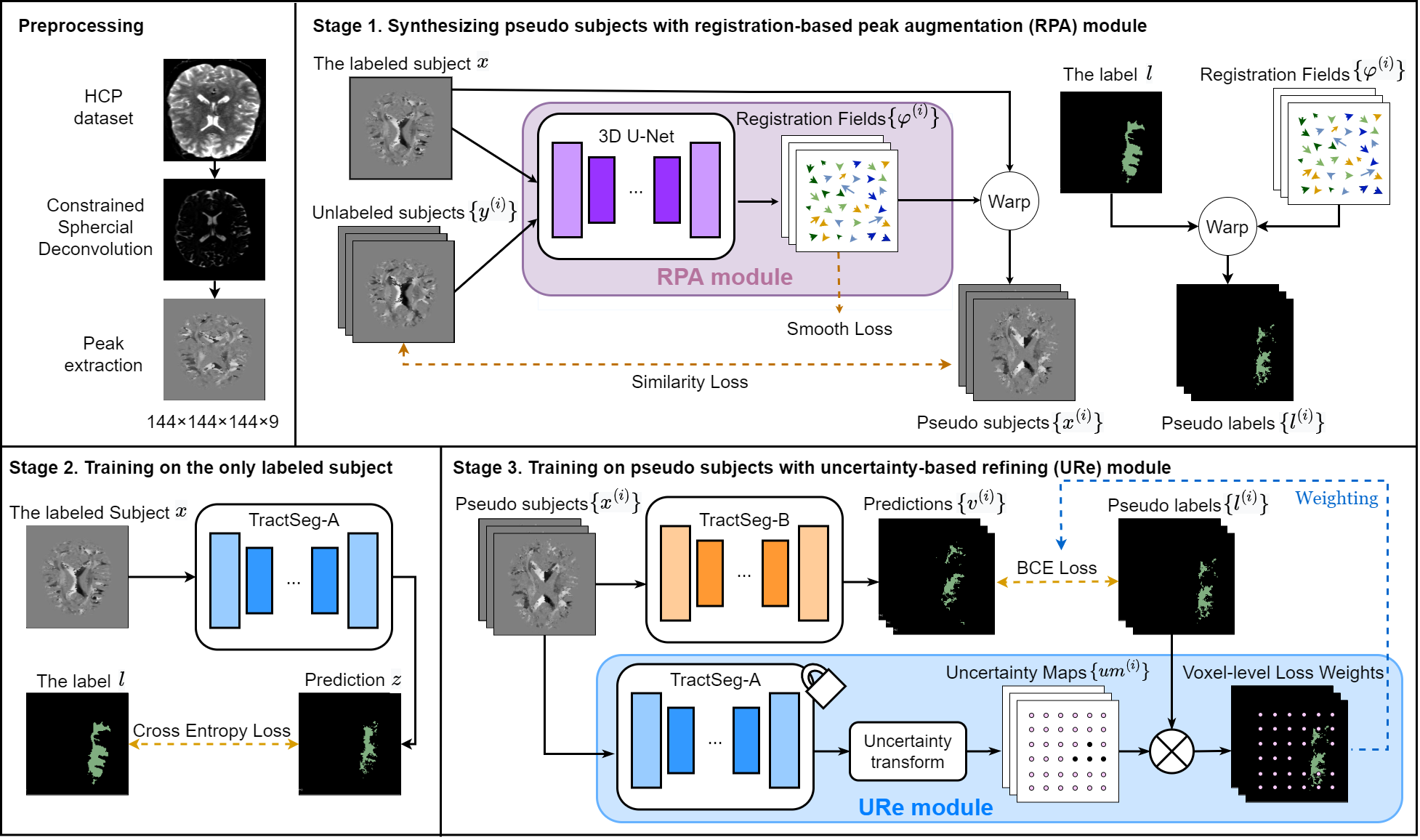}

\caption{Overview of registration- and uncertainty-based framework for WM tract segmentation with only one labeled subject.}

\label{fig:1}
\end{figure*}	

\section{MATERIALS AND METHODS}
\label{sec:2}

\subsection{dMRI Datasets and Data Preprocessing}
\label{sec:2.1}
We use the dataset from TractSeg \cite{Wasserthal2018-jj}, including 105 subjects from the Human Connectome Project (HCP) \cite{Van_Essen2013-ne}. Each subject has 72 tracts annotated by neuroanatomists. We follow the same dataset split as \cite{Wasserthal2018-jj}, using 84 subjects for training, and the rest 21 subjects for testing. Regarding our subject-level one-shot segmentation setting, we only use one annotated subject out of the 84 training subjects. As shown in Fig.\ref{fig:1}, we use multi-shell multi-tissue constrained spherical deconvolution (CSD) method \cite{Tournier2007-te} with all gradient directions to transform dMRI data to fiber orientation distribution function (fODF) peaks data, which has 9 channels corresponding to sagittal, axial, and coronal directions (each direction has three channels), as in \cite{Wasserthal2018-jj}. Compared with raw dMRI data, the networks trained with fODF peaks data achieve better tract segmentation results \cite{Wasserthal2018-jj}. Therefore, we directly segment tracts in the field of fODF peaks, and all subjects mentioned in the following are in this field.

\subsection{Registration- and Uncertainty-Based Framework for Subject-Level One-Shot White Matter Tract Segmentation}
As shown in Fig.\ref{fig:1}, we propose a registration- and uncertainty-based framework for subject-level one-shot tract segmentation. Three stages of our proposed framework are as below.

\noindent\textbf{Stage 1.}
Our proposed RPA module uses deep-learning-based registration \cite{Balakrishnan2019-zj,Zhao2019-af,Zhang2022-xr} to obtain pseudo labels for unlabeled subjects in our subject-level one-shot tract segmentation task. To alleviate the scarcity of labeled peak subjects, studies usually use the traditional data augmentation (e.g., rotation, flipping, and cropping) \cite{Wasserthal2018-jj} or linear registration-based augmentation \cite{Lu2021-fp}. Our RPA module based on more advanced deep-learning-based registration methods \cite{Balakrishnan2019-zj,Zhao2019-af,Zhang2022-xr} can potentially obtain the high-quality pseudo dataset. We modified registration methods in \cite{Balakrishnan2019-zj,Zhao2019-af}, which are originally for MRI data, to be used for our peak subjects. 
Let ${\{x,l\}}$ be the only labeled subject and its corresponding tract segmentation label, and ${y^{(i)}}$ be a set of unlabeled subjects on a spatial domain $R^3$. Pseudo peak subjects and their corresponding labels are synthesized with the RPA module. Specifically, a spatial transform model is learned to register the labeled subject to unlabeled subjects. Pseudo subjects and corresponding pseudo labels are generated by deep-learning-based registration using 3D U-Net \cite{Ronneberger2015-eu}. Set $g(x,y^{(i)})=u^{(i)}$ is the spatial transform model, where $\theta$ are the model parameters, and the output of the model $u$ is the voxel-wise displacement field. The deformation function $\varphi^{(i)} = id + u^{(i)}$, where $id$ is the identity transform \cite{Balakrishnan2019-zj}. Therefore, the pseudo subject $x^{(i)}$ and the corresponding pseudo label $l^{(i)}$ are as followed:
\begin{equation}
	x^{(i)}= x \circ \varphi^{(i)}, 
	\label{equation:1}
\end{equation}
\begin{equation}
	l^{(i)}=l \circ \varphi^{(i)}.
	\label{equation:2}
\end{equation}
For each voxel $p\in \Omega$, smooth loss $L_{smooth}$ and similarity loss $L_{sim}$ are considered as the registration loss: 
\begin{equation}
	L_{smooth}(\varphi^{(i)})= \sum_{p \in \Omega }^{} 
	\left \| \nabla u(p)  \right \| ^{2} ,
	\label{equation:3}
\end{equation}
\begin{equation}
	L_{sim}(x^{(i)},y^{(i)})=\frac{1}{\left | \Omega  \right | } \sum_{p \in \Omega}^{} \left [ y^{(i)}(p)-x^{(i)}(p) \right ]^{2} .
	\label{equation:4}
\end{equation}
$L_{smooth}$ is used to penalize the spatial variations in $\varphi^{(i)}$, and $ L_{sim}$ is used to penalize the difference between pseudo subject $x^{(i)}$ and unlabeled subject $y^{(i)}$. We balance $L_{smooth}$ and $L_{sim}$ with hyperparameter $\gamma$:
\begin{equation}
	L_{reg}=L_{smooth}+\gamma L_{sim}.
	\label{5}
\end{equation}

\noindent\textbf{Stage 2.}  
To further improve the quality of the pseudo dataset from Stage 1, we train a TractSeg network (TractSeg-A) on the only labeled subject to enable the tract segmentation and evaluate the quality of pseudo labels in the voxel-level. The trained TractSeg-A is used to calculate the voxel-level uncertainty map (Stage 3) to refine the pseudo dataset. 

TractSeg decomposes a 3D subject into 2D slices in three planes (sagittal, axial, and coronal planes) and trains a 2D U-net network with them. During inference/testing, in each plane, TractSeg reassembles the 2D slices of the prediction into a 3D subject prediction. The mean value of tract segmentation prediction of three planes is used as the segmentation prediction result of this subject. Set $z=m(x)$ is the prediction of TractSeg. We use binary cross entropy loss as the loss function of TractSeg:
\begin{equation}
	\begin{split}
		L_{u}= -\frac{1}{n}  \sum_{j=0}^{n} 
		\left [ l\cdot \log_{}{m(x)} + (1-l) \log_{}{(1-m(x))} \right ],  
		\label{equation:6}
	\end{split}
\end{equation}
where $n$ is the number of tract classes. 

\noindent\textbf{Stage 3.}
We use pseudo subjects and labels $\{x(i),l(i)\}$ that are refined using the proposed URe module to train TractSeg-B for predicting final results. The URe module improves the quality of pseudo subjects and labels by filtering out voxels that are not trustworthy using voxel-level uncertainty maps generated from TractSeg-A. First, 2D slices of pseudo subjects $x(i)$ are input into TractSeg-B for training. Set the TractSeg-B as $t(x(i))=v(i)$, where $v(i)$ is the segmentation prediction. Similar to Eq. \ref{equation:6}, loss of initial segmentation is as followed:
\begin{equation}
	\begin{split}
		L_{pseudo}=-\frac{1}{n}  \sum_{j=0}^{n} 
		[ l\cdot \log_{}{m(x)} + (1-l) \log_{}{(1-m(x))} ]. 
		\label{equation:7}
	\end{split}
\end{equation}
At the same time, to quantify the quality of pseudo subjects on voxel-level, $x(i)$ is input into parameter-frozen TractSeg-A. When the prediction value (from binary-cross-entropy loss) of TractSeg-A is closer to 0 or 1, it means that the prediction confidence of this voxel is higher. Based on that, we set the output of TractSeg-A to be $z(i)$, which is transformed into an uncertainty map $um^{(i)}$ through the uncertainty transform (URe module):
\begin{equation}
	um^{(i)}=\left\{\begin{matrix}
		2 \cdot  z^{(i)} -1,& if \quad z^{(i)}>0.5,\\
		1-  2 \cdot  z^{(i)},& otherwise.
	\end{matrix}\right.
	\label{equation:8}
\end{equation}
Finally, $um^{(i)}$ is used to weight the initial segmentation loss:
\begin{equation}
	L_{weight} = um^{(i)} \odot L_{pseudo}.
	\label{equation:9}
\end{equation}
$L_{weight}$ reduces the weight of prediction from untrustworthy voxels (voxels with lower prediction confidence).

\begin{figure*}[!h]
	\centering\includegraphics[width=16.5cm]{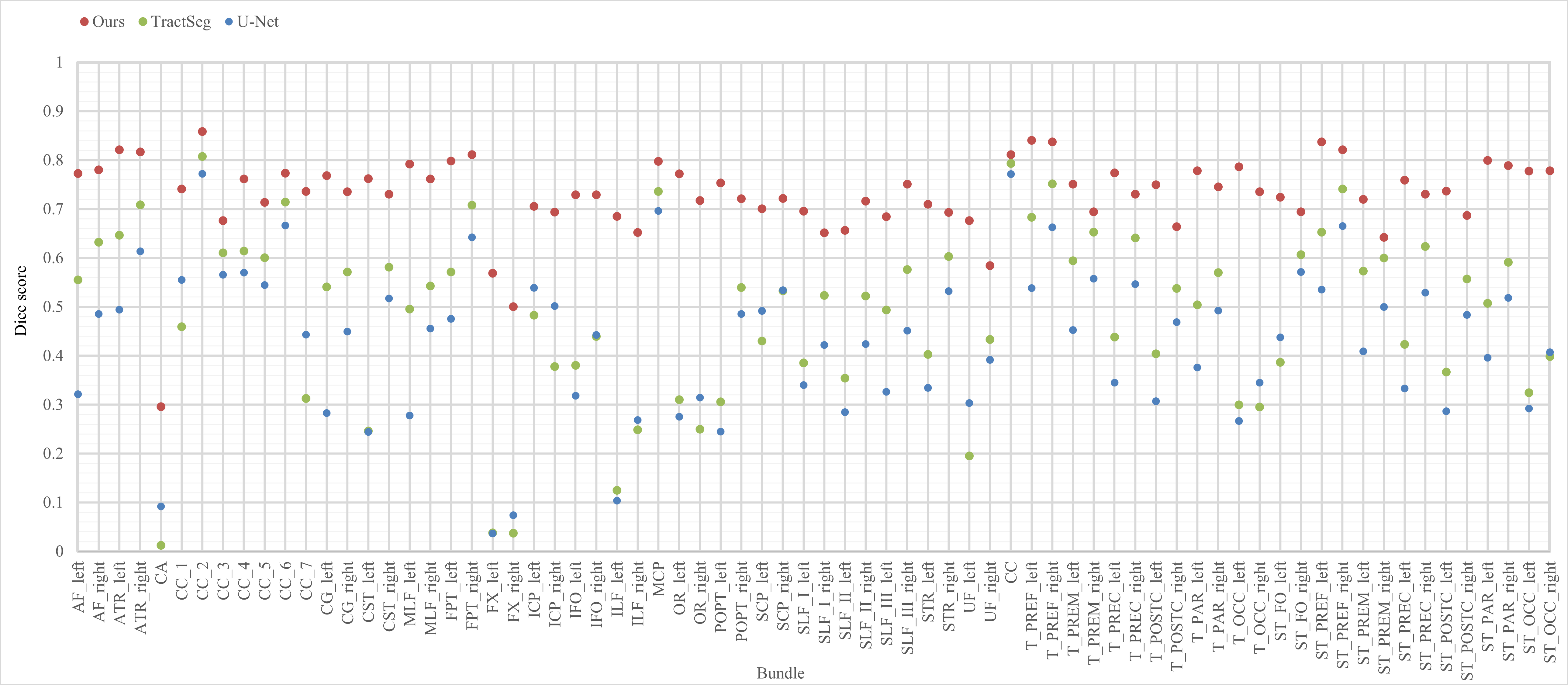}
	\caption{The mean Dice scores of all 72 tracts on the test set for our proposed method and compared methods.}
	\label{fig:2}
\end{figure*}

\section{EXPERIMENTS AND RESULTS}
\label{sec:3}
\subsection{Implementation Details}

\noindent\textbf{Training.}
In Stage 1, we use fODF peaks, which have a size of $144\times 144\times 144\times 9$, as the input of the 3D U-Net \cite{Ronneberger2015-eu}. The size of synthesized pseudo subjects is the same as input. The 3D U-Net is trained with an Adam optimizer. The learning rate is 0.001, the batch size is 1, the epoch is 100, and the hyperparameter $\gamma$ is 0.02. Hyperparameters in Stage 1 are referenced from \cite{Zhao2019-af} and tuned on our dataset.
Since there are 83 unlabeled subjects in the training set and a pseudo subject is generated for each unlabeled subject, 83 pseudo subjects are generated for subsequent training.
In Stage 2, the input of the TractSeg-A is a 2D image (slice) with a size of $144\times 144\times 3$ (sagittal, axial, and coronal planes). The output is the segmentation result of 72 tracts in the corresponding plane, and the size of the output is $144\times 144\times 72$. The TractSeg-A is trained with a learning rate of 0.02 and Adamax optimizer. The batch size is 48, the epoch is 200, and the dropout rate is 0.4. In Stage 3, parameters of TractSeg-A are frozen, and the TractSeg-B has the same input size, output size, and hyperparameters as the TractSeg-A from Stage 2. Hyperparameters in Stage 2 and 3 are referenced from \cite{Wasserthal2018-jj} and tuned on our dataset.

\noindent\textbf{Testing.}
During testing, we only use the trained TractSeg-B to get the final tract segmentation result. We stack 2D slice predictions to get 3D predictions (size of $144\times 144\times 144\times 72$) for the whole brain. The mean predictions of three planes (sagittal, axial, and coronal) are calculated as the final prediction results.

The above training and testing are performed with Pytorch (v1.10) on a NVIDIA GeForce RTX 3090 GPU machine.
\subsection{Comparison Experiments and Ablation Study}
We perform comparison experiments and ablation studies on the HCP test set. Our performance evaluation is based on the widely used metric, Dice score \cite{Wasserthal2018-jj,Lu2022-iv,Liu2022-se}.

\noindent\textbf{Overall Quantitative Comparison Experiments.}
We compare our method to U-Net, a popular deep-learning-based segmentation method, and TractSeg, a state-of-the-art (SOTA) white matter tract segmentation method, as shown in Table \ref{tab:1}. In our implementation, we train U-Net and Tractseg with only one annotated subject. Here, the difference between U-Net and TractSeg is that U-Net decomposes a 3D subject into 2D slices in only the sagittal plane for training and prediction, while TractSeg decomposes a 3D subject in three planes and calculates the mean value of tract segmentation prediction as the final result (as in \cite{Wasserthal2018-jj}). \textit{Ours (RPA+URe)} achieves a margin of 29.82\% and 24.16\% higher mean Dice score over TractSeg and U-Net, respectively.
\begin{figure}[h]
	\centering\includegraphics[width=8cm]{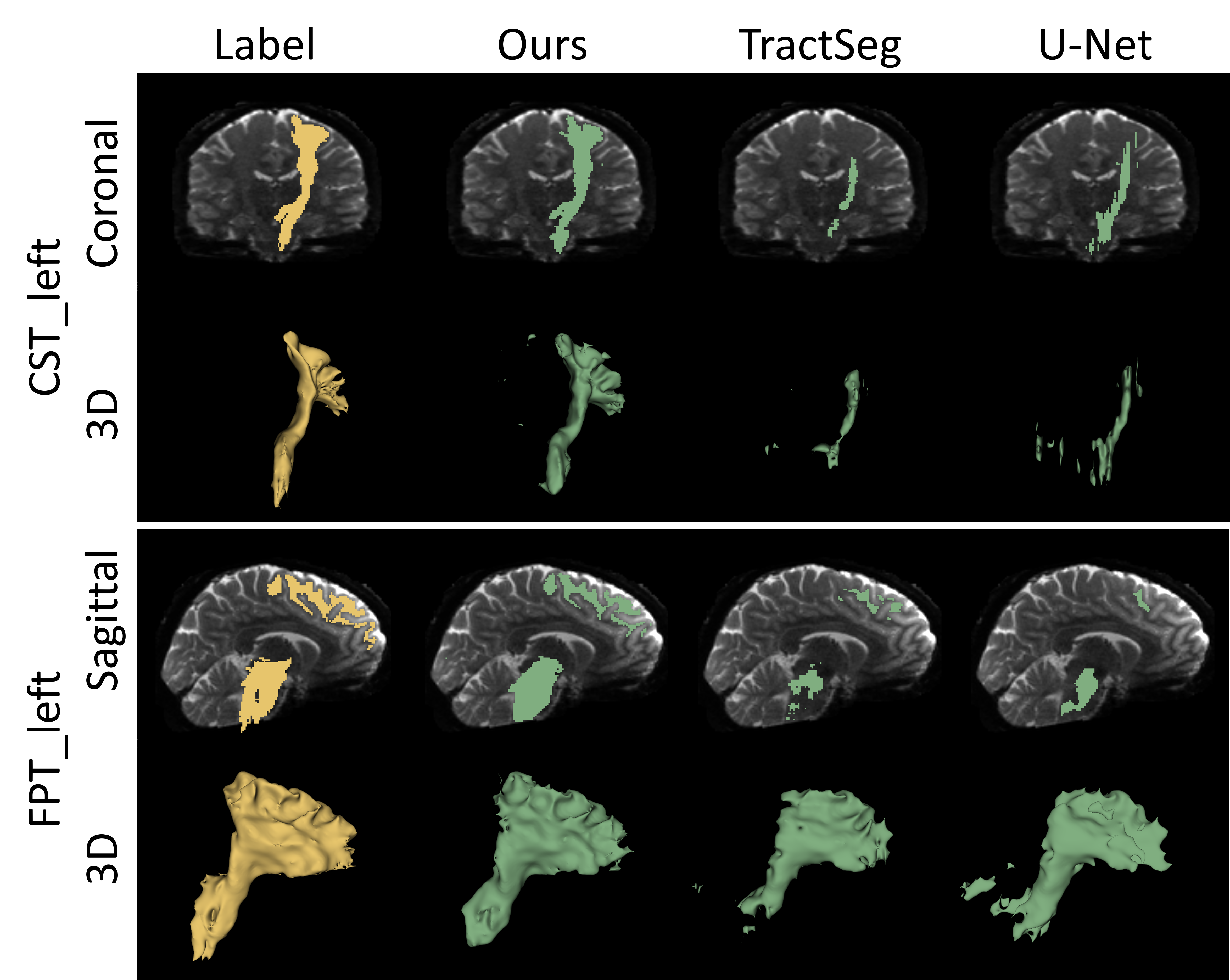}
	\caption{Visualization of tract segmentation results of two example tracts: left corticospinal tract (CST) and left fronto-pontine tract (FPT) on one subject. The yellow regions are labels, and the green regions are segmentation results of our method and compared methods.}
	\label{fig:3}
\end{figure}
\begin{table}[htbp]
	\centering
	\caption{Quantitative comparisons on HCP test set.}
	\begin{tabular}{|c|c|c|} 
		\hline
		& Methods    & Dice score             \\ 
		\hline
		\multirow{2}{*}{Comparison} & U-net      & 43.19±15.20\%          \\ 
		\cline{2-3}
		& TractSeg  & 48.85±17.58\%          \\ 
		\hline
		Ablation Study              & Ours (RPA) & 69.45±9.53\%           \\ 
		\hline
		\multicolumn{2}{|c|}{Ours (RPA+URe)}   & \textbf{73.01±8.14\%}  \\
		\hline
	\end{tabular}
 \label{tab:1}
\end{table}

\noindent\textbf{Ablation Study.}
We evaluate the impact of RPA and URe modules on tract segmentation performance, as shown in Table \ref{tab:1}. We design our framework based on TractSeg, therefore our method without RPA and URe modules (\textit{Ours (w/o RPA+URe)}) is the same as TractSeg. \textit{Ours (RPA+URe)} achieves 73.01\% mean Dice score, outperforming \textit{Ours (RPA)} and \textit{Ours (w/o RPA+URe)} by 3.56\% and 24.16\%, respectively. \textit{Ours (RPA+URe)} also achieves the lowest standard deviation of dice score compared with other methods. These results demonstrate the effectiveness of novel RPA and URe modules in our framework for subject-level one-shot tract segmentation.

\noindent\textbf{Quantitative Result on Every Tract.}
Fig. \ref{fig:2} shows the mean Dice scores of all 72 tracts on the test set for our proposed method and compared methods. The full name of each tract can be seen in \cite{Wasserthal2018-jj}. Compared with two SOTA methods, our method has the highest mean Dice score on all 72 tracts.

\noindent\textbf{Visualization of Tract Segmentation Results.}
 In Fig. \ref{fig:3}, we show the visualization of tract segmentation results for different methods. We observe that \textit{ours (RPA+URe)} can generate more complete and accurate segmentation results compared with other methods, even when the tract is very thin.

\section{CONCLUSION}
\label{sec:4}
In this work, we proposed a novel registration- and uncertainty-based framework for subject-level one-shot WM tract segmentation. Our method leveraged the proposed RPA module to synthesize pseudo subjects and their corresponding labels, and the proposed URe module for refining the low-confidence voxels in the synthesized subjects. Comparison results show that our method outperformed other SOTA methods and its ablated version by a large margin. Overall, our method achieved accurate tract segmentation of the whole brain using only one labeled subject.
\section{COMPLIANCE WITH ETHICAL STANDARDS}
\label{sec:5}
This research study was conducted retrospectively using human subject data made available in open access by Human Connectome Project \cite{Van_Essen2013-ne}. Ethical approval was not required.



%

\bibliographystyle{IEEEbib}
\bibliography{References}

\end{document}